\documentclass[11pt]{article}
\usepackage[margin=1in]{geometry}
\usepackage{amsfonts}
\usepackage{amsmath,amsthm,amssymb}
\usepackage{latexsym}
\usepackage{epic}
\usepackage{epsfig}
\usepackage{hyperref}
\usepackage{verbatim}
\usepackage[justification=centering]{caption}
\usepackage{enumitem}

\usepackage{amsfonts}
\usepackage{latexsym}
\usepackage{epic}
\usepackage{epsfig}
\usepackage{verbatim}
\usepackage{enumitem}
\usepackage{algorithmic}
\usepackage{setspace}
\usepackage{thmtools, thm-restate}
\renewcommand\thmcontinues[1]{(repeating \ref{#1} on p. \pageref{#1})}
\newtheorem{theorem}{Theorem}[section]
\newtheorem{corollary}[theorem]{Corollary}
\newtheorem{lemma}[theorem]{Lemma}
\newtheorem{proposition}[theorem]{Proposition}

\newtheorem{definition}[theorem]{Definition}

\newtheorem{fact}[theorem]{Fact}

\newcommand{\cross}{\times}

\newcommand{\set}[1]{\left\{ #1 \right\}}
\newcommand{\union}{\cup}
\newcommand{\intersect}{\cap}

\newcommand{\sm}{\setminus}

\renewcommand{\tilde}{\widetilde}
\renewcommand{\bar}{\overline}

\DeclareMathOperator{\poly}{poly}

\def\max{\qopname\relax n{max}}

\def\argmax{\qopname\relax n{argmax}}
\def\avg{\qopname\relax n{avg}}

\def\Pr{\qopname\relax n{\mathbf{Pr}}}
\def\Ex{\qopname\relax n{\mathbf{E}}}

\newcommand{\RR}{\mathbb{R}}
\newcommand{\RRp}{\RR_+}

\def\B{\mathcal{B}}

\def\D{\mathcal{D}}

\def\G{\mathcal{G}}

\def\L{\mathcal{L}}
\def\M{\mathcal{M}}

\def\R{\mathcal{R}}
\def\S{\mathcal{S}}

\def\V{\mathcal{V}}

\def\eps{\epsilon}
\def\sse{\subseteq}

\newcommand{\eat}[1]{}

\newenvironment{lp*}{\begin{equation*}  \begin{array}{lll}}{\end{array}\end{equation*}}

\DeclareMathOperator{\Rev}{Rev}

\begin{document}


\title{Sampling and Representation Complexity of Revenue Maximization}
 \author{
 Shaddin Dughmi \\ University of Southern California \\ {\tt shaddin@usc.edu} \and
 Li Han \\ University of Southern California \\ {\tt han554@usc.edu} \and
 Noam Nisan \\ Hebrew University and Microsoft Research \\ {\tt noam@cs.huji.ac.il}
 }

\maketitle
\begin{abstract}
We consider (approximate) revenue maximization in auctions where the distribution on input valuations is given via ``black box'' access
to samples from the distribution.  We observe that the number of samples required -- the sample complexity -- is tightly related to the 
representation complexity
of an approximately revenue-maximizing auction.  Our main results are upper bounds and an exponential lower bound on these
complexities.
\end{abstract}








\maketitle




\section{Introduction}

This paper studies revenue maximization in multi-parameter auctions in a Bayesian setting, a question that has received 
much attention lately.  
To place our results in context,
let us start with a high-level overview.  

\subsection{Background}

In the general (quasi-linear, independent-private-value, Bayesian) setting, 
we have a mechanism designer that can choose an {\em outcome}
from a set $A$ of possible ones.  There are $n$ {\em bidders}, each of which has a {\em valuation}
$v_i : A \rightarrow \Re$ that gives a real value for each possible outcome. These valuations are private, and
the mechanism designer only knows that each $v_i$ comes from a {\em prior distribution} $\D_i$ on valuations.  Based on these
distributions, the mechanism designer must design a {\em mechanism} that determines, for each profile of bidder valuations,
an outcome which may be probabilistic -- a {\em lottery} -- 
and a {\em payment} 
from each player.  
The rational behavior of the bidders is captured by two sets of constraints:
{\em Incentive constraints} require that no bidder can improve his expected utility by behaving according 
to -- i.e. ``reporting to the mechanism'' --
another valuation $v'_i$.  {\em Individual Rationality constraints} require that bidders never lose
from participating in the mechanism.  Under these two sets of constraints the mechanism designer's goal is to maximize his expected 
{\em revenue}. 

Myerson's classical work \cite{Myerson81} completely solves the problem for the special case of {\em single parameter} valuations,
where each $v_i$ is effectively captured by a scalar.  In this case, the optimal mechanism has a very simple form: first a
simple transformation is performed on each $v_i$ separately (in a way that depends on $D_i$ but not the other players' valuations).
Then the deterministic outcome that maximizes the sum of the transformed (``virtual'') valuations is chosen, and the payments
are then given by a simple critical value rule.  It turns out that this completely breaks down once we leave the
single-parameter settings and it is known that deterministic mechanisms can be significantly inferior to general
ones that are allowed to allocate lotteries \cite{MM88, Tha04, MV06, pricerand, chawla10b}, and that good revenue may require complex 
mechanisms \cite{hart-nisan-b}.  Moreover, this is so even in very simple settings such as auctions with a single bidder and two items.

It is well known that even in the general multi-parameter case, the revenue maximizing mechanism is obtained using linear optimization: 
the variables are the probabilities of each outcome
for each profile of valuations and the payments of each player for each profile of valuations, and the constraints are
the incentive constraints and the individual rationality constraints that turn out to be linear in these variables. While this 
may seem very encouraging both from a characterization and from a computation point of view, there is a rub: this linear programming
formulation hides several exponential blowups in the natural parameters in most settings.  Two types of such blowups, and how to overcome 
them, have received considerable attention recently: an exponential blowup in $n$, the number of bidders, that is a result of 
the fact that we have variables for each {\em profile} of valuations (e.g. \cite{alaeibayesian, alaeibayesian2, CDW12, huangcai}), and the fact that, in the 
case of various multi-item auctions, $m$,
the size of the outcome space, is naturally exponential in the number of items (e.g. \cite{chawla10a, DW12, CDW12b}). 

\subsection{Our Setting}

In this paper we study a third type of exponential blowup that was not yet considered: the size of the support of
each $\D_i$ -- the size of the valuation space -- is naturally exponential in the number of outcomes $m$.  Formally, it is often
a continuum since a valuation assigns a real value to each alternative, but even once we discretize 
(as we certainly will have to do for any computational purpose), then the valuation space has size that is 
exponential in $m$.  Do we need to pay this exponential price in $m$ or perhaps one can optimize revenue -- at least
approximately -- in time polynomial in the number of outcomes?

For most bite of our main result, which is negative, we focus on the simplest scenario of this type, one that exhibits only this
exponential blowup in the size of the outcome space, and no others.  Specifically, we will have a {\em single bidder} who
is bidding for one of the $m$ abstract alternatives in $A$.  For a single bidder, this generic mechanism design setting is 
essentially equivalent\footnote{The formal distinction is that in the unit-demand setting we may allow to not allocate any item 
to the bidder, which in the generic mechanism design setting would correspond to an 
additional possible outcome ``*'' that has a fixed value $v(*)=0$.} 
to an auction setting studied e.g. in \cite{chawla10b, pricerand} in which $A$ is a set of items for sale, a 
``unit demand'' bidder is interested in acquiring at most a single item and has a potentially different 
value $v(j)$ for each item $j \in A$. Furthermore, for a single bidder,
an auction is just a pricing scheme, giving a {\em menu} that assigns a price for each possible lottery $(x_1 ... x_m)$ where
$x_j \ge 0$ is the probability of getting item $j$ and $\sum_{j \in A} x_j \le 1$.  

As observed in
\cite{pricerand}, if $\D$ (the prior distribution over $v$) 
happens to have a ``small'' support then the linear program is small and we are done.
However, this is typically not the case: even if we restrict all item
values to be $0$ or $1$ there are $2^m$ possible valuations and the linear program is exponential. So how can we even represent
$\D$ in such a case?  In some cases $\D$ will have some simple structure, e.g. it may be a product distribution over item values, 
in which case it will have some succinct representation that depends on this structure.  In
other cases it may come from ``nature'' and we will, in some sense, have to ``learn'' $\D$ in order to construct our mechanism.  Both
of these scenarios can be captured by
a black box {\em sampling} model in which our access to $\D$ is by means of getting a sequence of
samples $v$, each one chosen independently at random from $\D$.  Will we be able to design an (approximately) revenue-maximizing
auction for $\D$ using a reasonable number of samples?  

The focus of this paper is the general case where $\D$ is not necessarily a product distribution.
The case of product distributions was previously studied by \cite{chawla07, chawla10b} who together achieve a polynomial-time constant-factor approximation
to the optimal revenue, and by \cite{ptas_indep} that gets a quasi-polynomial-time $(1+\epsilon)$ approximation to the 
revenue of {\em deterministic auctions}.  It is known, however, that
the general case of correlated distributions on item values is harder, e.g. deterministic prices can not provide a constant approximation \cite{pricerand}. 

The most natural approach for maximizing revenue from a distribution $\D$ that is given as a black box would be to sample some polynomial number of valuations from it, construct 
a revenue maximizing auction for this sample, and hope that the constructed auction also has good revenue on the original
distribution $\D$.  This, however, may fail terribly even for symmetric product distributions: 
Let $\D$ be the distribution where item values are chosen identically independently at random with the
following probabilities, where $\delta>0$ is some small constant: with probability $\delta$: $v(j)=1$; 
with probability $\delta/m$: $v(j)=2$; and otherwise: $v(j)=0$. 
Clearly just asking for a price of $1$
for each item will give revenue of very close to 1 (and somewhat higher revenue is possible, taking advantage of the $2$'s).  Yet, we show that
with high probability the following is an optimal auction for a sample drawn from this distribution: 
Price each individual item at $2$ and, for price $1$, offer lotteries only for the sets of $1$-valued 
items that were represented in the sample.  
None of these lotteries would be desirable
by a new random sample from $\D$, and thus this auction will 
obtain only $\Theta(\delta)$ revenue from $\D$.

The astute reader will recognize this failure as a classic case of over-fitting: the optimal mechanism for the sample
is so specifically targeted
to the sample that it loses any optimality for the real distribution $\D$.  The remedy for such over-fitting is well known:
we need to ``discourage'' such tailoring and encourage ``simple'' auctions.  At this point we need to specify
what ``simple'' auctions are, a question that is also closely linked to the question of how  to 
{\em represent} the auction that we output.  The simplest answer is the ``menu-complexity'' suggested in
\cite{hart-nisan-b}: we measure the complexity as the number of possible allocations of the auction,
i.e. as the number of entries in the menu specifying the auction.  
More complex representations that may be more succinct for some auctions are also possible.  Our lower
bounds will apply to any {\em auction representation language}.  Formally, an auction representation language is an arbitrary 
function that
maps binary strings to auctions.  The {\em complexity of an auction} in a given language is just
the length of the smallest binary string that is mapped to this auction.  Thus for example the menu-size of an auction
corresponds to complexity in the representation language where the menu-entries are explicitly represented by listing
the probabilities and price for each entry separately.\footnote{Counting bits, 
we are off from just counting the number of menu entries
by a factor of $O(m \log \epsilon)$ where $\epsilon$ is the precision in which
real numbers are represented. We will focus on exponential versus polynomial complexities, so do not assign much importance to
this gap.}  

\subsection{Our Results}

Fixing an auction representation language,  we  present the following sample-and-optimize template 
for revenue maximization that
takes into account the complexity of the output auction (in said representation language).
A similar idea is used in the context of prior-free mechanism design by~\cite{balcan}.
In our setting, beyond the dependence on complexity of the auction, the number of samples 
needs to depend (polynomially) on three other parameters: the number of items $m$, the required
precision $\epsilon$, and the range of the values. Specifically, assume that all valuations in
the support of $\D$ lie in the bounded range $1 \le v(j) \le H$ for all $j$. Since most of the expected value of a valuation
may come from events whose probability is $O(1/H)$, it is clear that we will need $\Omega(H)$ samples to even notice these 
events.\footnote{This also explains why bounding the range of valuations, as we do, is required for the sampling question to make any sense.
Equivalently, we could have instead bounded the variance of $\D$ without significantly changing any of our results.}

\vspace{0.1in}
\noindent
{\bf Sample-and-Optimize Algorithm Template}
\begin{enumerate}
\item Sample $t=poly(C,m,\epsilon^{-1},H)$ samples from $\D$.
\item Find an auction of complexity at most $C$ that maximizes (as much as possible) 
the revenue for the uniform distribution over the sample and output it.
\end{enumerate}
\vspace{0.1in}

To convert this algorithm template to an algorithm, one must specify $C$ and then demonstrate an effective algorithm
to find the auction of complexity $C$ that achieves high revenue on the sample. (Note that without the complexity
bound, even fully maximizing
revenue is efficiently done using the basic linear program, but this does not carry-over to complexity-bounded maximization.)
Once the complexity is bounded, at least information-theoretically, the ``usual'' learning-like uniform convergence bounds indeed 
apply
and we have: 

\begin{proposition} \label{apx-C}
Fix any auction representation language, and take an algorithm that follows
this template and always produces an auction that approximates to within an $\alpha$ factor the optimal revenue from the {\em sample}
over all auctions of complexity $C$.
Then the produced auction also approximates to within a factor of $(1-\epsilon)\alpha$ the revenue from 
the real prior distribution $\D$
over all auctions of complexity $C$.
\end{proposition}


Thus we get approximate revenue maximization over all complexity-$C$ auctions. However,  if this limited class  is inferior to 
general auctions, this does not yield approximate revenue maximization over all auctions, which is our goal. 
The following questions thus remain:

\begin{enumerate}
\item What is the complexity $C$ required of an auction in order to obtain good revenue?  How good an approximation can we get
when we require $C$ to be polynomial in $H$ and $m$?
\item What is the computational complexity of step 2 of the algorithm template, 
i.e. of constructively finding a mechanism 
that maximizes revenue over mechanisms of bounded complexity $C$?   
\end{enumerate}

We provide definitive answers to the first question and preliminary answers to the second.
Apriori, it is not even clear that any finite complexity $C$ 
suffices for getting good revenue (for fixed $H$ and $m$). Previous work  
(\cite{DW12}) 
implies that arbitrarily good approximations are possible using menu-size 
complexity that is polynomial in $H$ and exponential in $m$  and for
the special case $m=2$ even poly-logarithmic size in $H$, \cite{hart-nisan-b}.  This
is done by taking 
the optimal auction and ``rounding''
its entries. This is trickier than it may seem since
a slight change of probabilities may cause a great change in revenue, so such proofs 
need to carefully adjust the rounding of probabilities and prices making sure that significant revenue is never 
lost.\footnote{Technically, there is no countable $\epsilon$-net of auctions in the sense of approximating 
the revenue for every distribution.  Instead, one needs to construct a ``one-sided'' net,
and to loose a quadratic factor in $\epsilon$ as well.}
We describe a general way to perform this adjustment, allowing us to tighten these results: we get
poly-logarithmic dependence in $H$ for general $m$, stronger bounds
for ``monotone'' valuations, and do it all effectively in a computational sense.  
Monotone valuations are restricted to have $v(1) \le v(2) \le ... \le v(m)$, and naturally model 
cases such as 
values for a sequence
of ad-slots or for increasing numbers of items in a multi-unit auction.

\begin{theorem}\label{thm:menusize}
For every distribution $\D$ and every $\epsilon>0$ there exists an auction with menu-size complexity at most 
$C=\left(\frac{\log H + \log m + \log\epsilon^{-1}}{\epsilon}\right)^{O(m)}$ whose revenue is at least $(1-\epsilon)$
fraction of the optimal revenue for $\D$.  
For the special case of distributions over ``monotone'' valuations, 
menu-size complexity of at most $C=m^{O((\log^3 H + \log^2
\epsilon)/\epsilon^2)}$ suffices.
Furthermore, in both cases these auctions can be computed in $poly(C,H,\epsilon^{-1},m)$ time by sampling $poly(C,H,\epsilon^{-1},m)$ valuations from $\D$.
\end{theorem}


So in general a menu of complexity exponential in $m$ suffices. A basic question is whether complexity  
polynomial in $m$ suffices (in terms of
menu-size or perhaps other stronger representation languages). 
The ``usual tricks'' suffice to show that an $O(\log H)$-approximation of the revenue is possible with small menus.

\begin{proposition} \label{pro:logH}
For every distribution $\D$ there exists an auction with $m$ menu-entries
(and with all numbers represented in $O(\log H)$ bits)
that extracts $\Omega(1/\log H)$ fraction of the optimal revenue 
from $\D$.  
Furthermore, this auction can be computed in polynomial time from $\poly(H)$ samples.
\end{proposition}

Can this be improved?  Can we get a constant factor approximation with polynomial-size complexity? Our main result
is negative.
Previous techniques that separate the revenue of simple auctions from that of general auctions do not suffice for proving 
an impossibility here for two reasons.  First, these bounds only apply to menu-size complexity and not to general auction representations; this is explicit in \cite{hart-nisan-b}
and implicit in \cite{pricerand}.\footnote{Since these papers
exhibit an explicit distribution that provides the separation, the optimal auction for this distribution can always
be specified in some language by just listing the few parameters of the distribution.}
Second, these bounds proceed by giving an upper bound to the revenue that a single menu-entry can extract.  Since small menus can extract an $O(1/\log H)$ fraction of the optimum revenue, 
such techniques can have no implications for sampling complexity since, as mentioned above,
$\Omega(H)$ is a trivial lower bound on the sampling complexity.  
Our main result shows that even for a small range of values $H$, auctions may need to be exponentially complex in $m$
in order to break the $O(1/\log H)$ revenue approximation ratio.  

\begin{theorem}\label{thm:kolmogorov}
For every auction representation language and every $1 < H < 2^{m/400}$ there exists a 
distribution $\D$ on $[1..H]^m$ such that every auction with complexity at most 
$2^{m/400}$ has revenue that is at most an $O(1/\log H)$ fraction of the optimal revenue for $\D$.
\end{theorem} 

Notice that this immediately implies a similar exponential lower bound on the number of samples needed: since we allow any
auction description language, simply listing the sample is one such language for which the lower bound holds.  

At this point we examine our second question, regarding the computational complexity of ``fitting'' an auction of 
low complexity to sampled data.  We show that
even given a menu-size bound, it is $NP$-hard to approximate the optimal revenue of auctions of that 
menu-size.\footnote{Here we would expect stronger auction representation languages to
only be harder to deal with.}

\begin{theorem}\label{thm:maxrev_nphard}
Given as input a sample of valuations and a target menu-size $C$, it is
NP-hard to approximate the optimal revenue achievable in size $C$ to within any factor 
better than $1-\frac{1}{e} \frac{H-1}{H}$.
\end{theorem}

This hardness result 
does not preclude a satisfactory answer to our original goal of effectively finding an auction that 
approximates the revenue also on the original distribution $\D$ since for that it suffices to find a 
``small'' auction with good revenue on the sample,  rather than the ``smallest'' one.  Thus a bi-criteria approximation 
to step 2 suffices\footnote{In fact, it is also necessary.}, and may be algorithmically easier: 
find a menu of size $\poly(C,m,H)$ which approximates (as well as possible) the revenue of the best 
menu of size $C$ over a given sample. 
Whether this bi-criteria problem can be solved in polynomial time is left as our first open problem.

Our second open problem concerns the question of structured distributions, specifically product distributions over item values
studied in \cite{chawla07,  chawla10b, ptas_indep}.  
Proposition \ref{apx-C} implies that, as these distributions can be succinctly represented,
polynomially many samples suffice for finding a nearly optimal auction
for product distributions.\footnote{This is directly 
implied when the item values have finite
(polynomial) support; the techniques used in section \ref{lot-auc} suffice for showing it in general.}  It is not clear, however,
how this can be done algorithmically and whether the simple menu-size auction description language suffices for 
succinctly representing
the (approximately) optimal auction.  Constant factor approximation with small menu-size (even deterministic menus) follow from \cite{chawla07, chawla10b}, but
a $(1+\epsilon)$-approximation is still open.



\section{Preliminaries}
\label{sec:prelim}

\subsection{The Model}
\label{sec:model}
In the \emph{single-buyer unit-demand mechanism design problem}, or the \emph{pricing problem} for short, we assume that there are   
$m$ ``items'' or ``outcomes'' $[m]=\set{1,\ldots,m}$, and a single risk-neutral
buyer equipped with a valuation $v \in \RRp^m$. Additionally, we assume the existence of an additional outcome $*$ for 
which a buyer has value $0$ -- e.g. the outcome in which the player receives no item.
We assume that $v$ is drawn from a distribution $\D$ supported on some  family of valuations $\V \sse \RRp^m$.  

We adopt the perspective of an auctioneer looking to sell the items in order to maximize his revenue.  After soliciting a bid $b \in \V$, 
the auctioneer chooses an \emph{allocation}, namely a (partial) lottery $x \in \Delta_m=\set{x \in \RRp^m : \sum_i x_i \leq 1}$ 
over the items, and a \emph{payment} $p \in \RRp$. Formally, the auctioneer's task is to design a \emph{mechanism}, 
equivalently, an auction, $(x,p)$, 
where $x: \V \to \Delta_m$ maps a player's reported valuation to a lottery on the $m$ items, and $p: \V \to \RRp$ maps the 
same report to a payment. When each allocation in the range of $x$  is a deterministic choice of an item, 
we say the mechanism (or auction) is \emph{deterministic}, otherwise it is \emph{randomized}.

To simplify our results, we usually assume that players' valuations lie in a bounded range. Specifically, we require that the 
support $\V$ of our distribution is contained in $[1,H]^m$, for some finite upper-bound $H$ which may depend on the 
number of items being sold. Given this assumption,  we restrict our attention without loss of generality to mechanisms 
with payment rules constrained to prices in $[1,H]\union\set{0}$. Moreover, we assume without loss of 
generality that $p(v) =0$ only if $x(p) = \vec{0}$.\footnote{An optimal mechanism satisfying these two 
properties always exists for all the problems we consider.}

Whereas our complexity results are independent of the representation of $\D$, our algorithmic 
results hold in the \emph{black-box model}, in which the auctioneer is given sample access to the 
distribution $\D$, and otherwise knows nothing about $\D$ besides its support $\V$. 


\subsection{Truthfulness and Menus}
We constrain our mechanism $(x,p)$ to be \emph{truthful}: i.e. bidding $b=v$ maximizes the buyer's 
utility $v \cdot x(b) - p(b)$. The well known characterization below reduces the design of such a 
mechanism to the design of a \emph{pricing menu}.

\begin{fact}[Taxation Principle]
A mechanism $(x, p)$ is truthful if and only if
there is a menu $M \subseteq \Delta_m \times \RR$ of allocation/price pairs such
that
\begin{equation}
(x(v), p(v)) \in \argmax_{(x, p) \in M} \{ v \cdot x - p \}
\end{equation} 
\end{fact}

We adopt the menu perspective through much of this paper, interchangeably referring to a mechanism (aka auction)
and its corresponding menu $M$. When interpreting a menu $M$ as a mechanism, we break ties in $v \cdot x - p$ in 
favor higher prices. When every allocation in the 
menu is a deterministic choice of an item, we call $M$ an \emph{item-pricing menu}, otherwise we call it a \emph{lottery-pricing menu}.

We also require our mechanisms to be \emph{individually rational}, in that the player's utility 
from participation in the mechanism is never negative. To enforce this, we assume that $(\vec{0},0)$ is in 
every menu. As described in Section \ref{sec:model}, we usually restrict valuations to $[1,H]^m$ and 
payments for non-zero lotteries to $[1,H]$. Therefore, we think of a menu as a subset of $\Delta_m \cross [1,H]$, and include $(0,0)$ implicitly.

\subsection{Auction Complexity and Benchmarks}
Given a mechanism $M$ and valuation $v$, we use $\Rev(M,v)$ to denote the payment of a buyer with valuation $v$ when participating in the mechanism. Given a distribution $\D$ over valuations, we use $\Rev(M,\D)= \Ex_{v \sim \D} \Rev(M,v)$ to denote the expected revenue generated by the mechanism when a player is drawn from distribution $\D$. We use $\Rev(\D)$ to denote the supremum, over all mechanisms $M$, of $\Rev(M,D)$. When $\M$ is a family of mechanisms, we use $\Rev(\M,\D)$ to denote $\sup_{M \in \M} \Rev(M,\D)$.

Recall that a auction description language is just a mapping from binary strings to mechanisms.  I.e. it is simply a way 
of encoding menus in binary strings.
The representation complexity of an auction in such a language is simply the length of the shortest string that is mapped to it.
For this paper, the only important property of auction description languages is that there are at most $2^C$ auctions
of complexity $C$.  (Of course, in applications we will also worry about its 
expressive power, its computational difficulty, etc.)
The simplest auction description language allows describing an auction by directly listing its menu entries one by one.  Each menu
entry is composed of $m+1$ numbers, and if all the numbers can be presented using $O(r)$ bits of precision, then the
total complexity of a $k$-entry auction in this format is $O(kmr)$.  In this paper we never need
more than $r=O(\log m + \log H + \log \epsilon^{-1})$ bits of precision, so the gap between menu-size (the number of menu entries) and
complexity using this language (the total number of bits used in such a description) is not significant.

We use $\Rev_k(\D)$ to denote the maximum revenue of a mechanism with complexity at most $k$. 
When using menu-size complexity, we say $M$ is a $k$-menu if $|M| \leq k$, and use $\M_k$ to denote the set of all k-menus, 
and $\M_\infty$ to denote the set of all menus.


\section{Sampling vs. Auction Complexity}

\subsection{Over Fitting with Complex Auctions}

In this subsection we will consider the basic sampling algorithm that makes a small number of samples and optimizes the auction for
this sample.  

\vspace{0.1in}
\noindent
{\bf Naive Sample-and-Optimize Algorithm}
\begin{enumerate}
\item Sample $t=poly(m)$ samples from $\D$.
\item Find an auction that maximizes 
the revenue for the uniform distribution over the sample and output it.
\end{enumerate}
\vspace{0.1in}

We will show that this does not work even for symmetric product distributions.

Let $\delta>0$ be some small constant and let $\D$ be the distribution on valuations where item 
values are chosen identically and independently at random as follows:
with probability $\delta$: $v(j)=1$; 
with probability $\delta/m$: $v(j)=2$; and otherwise: $v(j)=0$. 
We will show that optimizing
for a sample may give very low revenue on $\D$ itself:

\begin{proposition}
Take a random polynomial-size sample then, with high probability, there is an auction that is optimal for the sample and yet
its revenue from $\D$ is $O(\delta)$.
\end{proposition}

\begin{proof}
The following auction will get, for every valuation $v$ in the support, revenue that is exactly equal to the maximum item price in $v$, and thus
it will be optimal.  First we sell each item at price $2$.  For every $v$ in the sample {\em that does not value any item at $2$}, 
let $S_v$ denote the set of (about $\delta m$) items
that it gives value $1$ for, and our auction will offer a lottery that picks an item uniformly at random from $S_v$ for price $1$.

Let us say that a sample is typical if all valuations $v$ in it have between $\delta m /2$ and $2\delta m$ items with non-zero
value and additionally the intersection of any two such supports is at most of size $2\delta^2 m$.  We claim that our sample
is typical with very high probability: The expected size of the support of a random valuation from $\D$ is $\delta m$ and
taking two random valuations, their expected support intersection size is $\delta^2 m$.
Chernoff bounds show that we are far from this expectation with exponentially small probability, and this allows taking
the union bound over all polynomially-many pairs in the sample, to get the required properties simultaneously for the whole sample.

For every $v$ in our sample that does not value any item at $2$, we will surely get revenue $1$ for the uniform lottery on $S_v$, and this is as
high as possible.  Now the claim is that, for a typical sample, all $v$ in the sample that do have some item valued at $2$ will actually 
buy an item at price $2$ which again is highest possible.  
For this to be the case we need that $v$'s value from any lottery offered is less than $1$ (which would give him negative
utility from all menu entries that are priced at less than $2$).  This is so since his value from the uniform lottery on $S$ is at most 2 
times the probability that
$v$ gives non-negative value to 
a random element chosen from $S$, which by the assumption that the sample is typical is 
bounded from above by $(2\delta^2 m)/(\delta m/2) = 4 \delta$ for 
all $S$ that were offered
as lotteries.  We thus have that for typical samples this auction is optimal.

Now let us examine the revenue that this auction will get from $\D$: A random $v$ will have an item valued at $2$ with probability of about $\delta$, giving
us $O(\delta)$ revenue from these.  Otherwise, $v$ will buy a lottery over a set $S$ only if its value from it is $1$ 
which means that $v$'s support is fully contained in $S$.  For a random $v$ from $\D$ this happens, for any of the 
(polynomially many) lotteries in our auction,
only with exponentially small probability.
\end{proof}

\subsection{Uniform Convergence over Simple Auctions}

When we limit the ``complexity'' of the auction that our algorithm is allowed to produce
to be significantly smaller than the sample size, we can guarantee that the produced 
auction approximately maximizes revenue for the original distribution $\D$. 
The pertinent property here is that there cannot be too many auctions of low complexity.
This is all that it takes to prove the correctness of the basic sample-and-optimize algorithm template:

\vspace{0.1in}
\noindent
{\bf Sample-and-Optimize Algorithm Template}
\begin{enumerate}
\item Sample $t=poly(C,m,\epsilon^{-1},H)$ samples from $\D$.
\item Find an auction of complexity at most $C$ that maximizes (as much as possible) 
the revenue for the uniform distribution over the sample and output it.
\end{enumerate}
\vspace{0.1in}

\begin{proposition} (\ref{apx-C} from the Introduction)
Fix any auction representation language, and take an algorithm that follows
this template and always produces an auction $M$ of complexity at most $C$ (in the given language) 
that approximates, to within an $\alpha$ factor, the 
optimal revenue from the {\em sample},
over all auctions of complexity $C$.
Then, the produced auction also approximates, to within a factor of $(1-\epsilon)\alpha$, the revenue from 
the real prior distribution $\D$,
over all auctions of complexity $C$.
\end{proposition}

\begin{proof}
Take some auction $M$ that achieves some revenue $r \ge 1$ on $\D$.  By definition of revenue, $r=E_{v \sim \D}[p(v)]$.
In order to estimate it we take $t$ samples $v^1 ... v^t$ from $\D$ and we can directly apply Chernoff-Hoeffding bounds to estimate
the probability that the average revenue for our sample, $(\sum_i p(v^i))/t$, is far from its expectation.  The standard 
deviation of $\D$ is clearly bounded from above by $H$ so the probability that $|(\sum_i p(v^i))/t - r| > \epsilon/2$ is bounded
from above by $exp(-\Omega(t\epsilon^2 /H^2)$.  Taking $t=O((C + \log\delta^{-1})H^2/\epsilon^2)$ we get this probability to be bounded
from above by $\delta 2^{-C}$.

We now apply the union bound over all the at most $2^C$ auctions that have
complexity at most $C$, obtaining that with high probability (at least $1-\delta$)
for {\em every} auction with complexity at most $C$, its revenue over the sample is within $\epsilon/2$ of its revenue from $\D$.
In which case, an auction that comes to within a factor $\alpha$ of optimal for the sample is also within a factor of $(1-\epsilon)\alpha$
factor of optimal for the distribution (all over the auctions of complexity at most $C$.)
\end{proof}


\section{Constructions of Simple Approximating Auctions}

\subsection{From Rounding Lotteries to ``Rounding'' Auctions}\label{lot-auc}

When attempting to approximate a given auction (menu), it is quite natural to simply round all entries in the menu and hope that
this does not hurt the revenue significantly.  As mentioned in the introduction, this is not trivial since tiny decreases in
probabilities or tiny increases in price may be the ``last straw'' chasing away bidders that made knife's-edge choice of the 
entry.  This subsection
shows that, never the less, this may be done with a little further tweaking: once we have a good 
way to round lotteries we can ``round'' whole menus, loosing 
an approximation factor that is polynomially related to the rounding error.

\begin{definition}\label{def:lottery_cover}
Let $V$ be a set of valuations, and $L$ be a set of lotteries.  We say that $L$ $\epsilon$-covers $V$ if for every lottery
$x \in \Delta_m$ there exists a lottery $\tilde{x} \in L$ such that for every $v \in V$ we have that 
$x \cdot v \ge \tilde{x} \cdot v \ge (1-\epsilon) x \cdot v - \epsilon$. 
\end{definition}

\begin{lemma}\label{lem:lot-auc}
Let $L$ be a set of lotteries that $\epsilon$-covers a set of valuations $V$, then for every menu $M$ 
there exists a menu $\tilde{M}$ all of whose entries have lotteries in 
$L$ and have prices represented using $O(\log \epsilon^{-1} + \log H + \log m)$ bits such that for 
every $v \in V$, 
we have that $Rev(\tilde{M},v) \ge (1-\epsilon')Rev(M,v) - \epsilon'$, where $\epsilon'=O(\log H \sqrt{\epsilon})$.
Moreover, if the calculation of $\tilde{x}$ from $x$ is efficient then
so is the calculation of $\tilde{M}$ from $M$.
\end{lemma}

\begin{proof}

Define the rounding map $\xi: \Delta_m \to L$ so that $\tilde{x} = \xi(x)$ is as in Definition \ref{def:lottery_cover}. Let us partition the entries $(x,p)$ of the menu $M$ into levels, where each level $k$, for $k = 1 ... K$ with $K=O(\log H/\delta)$,
contains all entries whose price is in the range $(1+\delta)^{k-1} < p \le (1+\delta)^k$.
For every entry $(x,p)$ of $M$ in level $k$ we will put an entry $(x',p')$ in $\tilde{M}$ where
$x' = \xi((1-\epsilon)^{K-k}x)$ and $p'$ is obtained by first multiplying $p$ by a factor of
$(1-\epsilon)^K$, then rounding $p$ down to an integer multiple of $\epsilon$,
and then subtracting $2k\epsilon$.
Thus entries with lower level payments have their allocations reduced more than, and their payments reduced less than, 
those with higher payments.  Specifically notice that if $(x,p)$ is at level $k$ and $(y,q)$ is
at a lower level $k'<k$ then (A)
$p'-q' < ((1-\epsilon)^K p - 2k\epsilon) - ((1-\epsilon)^K q - 2k'\epsilon - \epsilon) \le 
(1-\epsilon)^K(p - q) - \epsilon$
and that (B) for every $v \in V$ we have that
$x' \cdot v-y' \cdot v = \xi((1-\epsilon)^{K-k}x) \cdot v - \xi((1-\epsilon)^{K-k'}y) \cdot v >
((1-\epsilon)^{K-k+1}x \cdot v - \epsilon) - (1-\epsilon)^{K-k'}y \cdot v \ge (1-\epsilon)^K (x \cdot v-y \cdot v) - \epsilon$.

Now take a valuation $v$ that, in $M$, chooses entry $(x,p)$ and thus $x \cdot v - p \ge y \cdot v - q$ for all
other entries $(y,q)$ in the menu and thus (C) $p - q \le x \cdot v - y \cdot v$.  In $\tilde{M}$ the valuation $v$ will choose 
some entry $(y',q')$.  We claim that $q$ cannot be at a lower level than $p$  since otherwise we have
$y' \cdot v - q' \ge x' \cdot v - p'$ and thus $x' \cdot v - y' \cdot v \le p' -q'$ which is impossible
since we get a contradiction combining it with (A) $p'-q' < (1-\epsilon)^K(p - q) - \epsilon$, (C) $p - q \le x \cdot v - y \cdot v$,
and (B) $(1-\epsilon)^K (x \cdot v-y \cdot v) - \epsilon < x' \cdot v-y' \cdot v$.

Let us now see how large a loss may we have incurred between payment $p$ and payment $q'$.  First, $q$ may be smaller
by a factor $(1+\delta)$ than $p$, then to obtain $q'$ we first lost a multiplicative factor of $(1-\epsilon)^K$ and then
an additive factor of at most $(2K+1)\epsilon$ (including the rounding).  Thus
$q' \ge (1-\delta)(1-\epsilon)^K p - (2K+1)\epsilon$.  Choosing $\delta = \theta(\sqrt{\epsilon})$ we have 
$K = \theta(\log H /\sqrt{\epsilon})$ and so $q' \ge (1-\epsilon')p - \epsilon'$ for $\epsilon' = O(\log H \sqrt{\epsilon})$.
\end{proof}

\noindent
{\bf Comment:} This proof also allows us leeway to round the valuations $v$.  Specifically, if one takes a distribution $\D$
with continuous support, and defines a ``rounded distribution'' $\D'$ obtained by taking samples from $\D$ and rounding
all values to integer multiples of $\epsilon$, then while an arbitrary auction $M$ that maximizes revenue for $\D'$ may give bad results
for $\D$, the ``rounded'' mechanism $\tilde{M}$ will also give high revenue for $\D$.

\subsection{Approximations for General Valuations}
So at this point we know that we just need to worry about rounding lotteries.  Once we round all values to a small
number of discrete values, we will get a small number of lotteries. Unfortunately, we need to use both an additive and
multiplicative approximation error: the multiplicative approximation error allows a 
large additive error when values are close to $H$; and the additive error saves us from having to approximate multiplicatively very small
probabilities.  Combining these two notions of error allows us to make do with $O(\log H / \epsilon)$ discrete levels of approximation.

\begin{proposition}
Let $R_\epsilon$ be the set of real numbers containing zero and all integer powers of $(1-\epsilon)$ in the range $[\epsilon/(Hm),1]$,
and let $L_\epsilon$ be the set of lotteries all of whose entries are in $R_\epsilon$.  Then $L_\epsilon$ $\epsilon$-covers the set of
all valuations $v \in [1,H]^m$.  Moreover, calculating $\tilde{x}$ from $x$ can
be done efficiently.
\end{proposition}

\begin{proof}
Let $\tilde{x}$ be obtained by rounding down each entry of $x$ to a value in $\R_\epsilon$.  
We have $\tilde{x}\cdot v \ge \sum_{i| x_i \ge \epsilon/(mH)} \tilde{x}_i v_i$ and since
for $x_i \ge \epsilon/(mH)$ we have
$\tilde{x}_i \ge (1-\epsilon) x_i$, we get 
$\tilde{x}\cdot v \ge (1-\epsilon) \sum_{i| x_i \ge \epsilon/(mH)} x_i v_i$.  To complete
the proof note that 
$x\cdot v\le\sum_{i|x_i\ge\epsilon/(mH)}x_i v_i + H\sum_{i|x_i<\epsilon/(mH)}x_i\le\sum_{i|x_i\ge\epsilon/(mH)}x_i v_i +\epsilon$.
\end{proof}

\begin{corollary}
\label{cor:cover_size_li}
There exists an $\epsilon$-cover of the set of all valuations $v:\{1...m\} \rightarrow [1,H]$ 
whose size is $((\log m + \log H + \log \epsilon^{-1})/\epsilon)^m$.
\end{corollary}

\begin{corollary} \label{partI} (Part I of theorem \ref{thm:menusize} from the introduction)
For every distribution $\D$ on $[1,H]^m$ and every $\epsilon>0$ there exists an auction with menu-size complexity at most 
$C=\left(\frac{\log H + \log m + \log \epsilon^{-1}}{\epsilon} \right)^{O(m)}$ whose revenue is at least $(1-\epsilon)$
fraction of the optimal revenue for $\D$.  
Furthermore, this auction can be computed effectively (in polynomial time in its size) from a sample of 
size $\poly(C,m,H,\epsilon^{-1})$ .
\end{corollary}

\begin{proof}
Combining corollary \ref{cor:cover_size_li} using $O((\epsilon/\log H)^2)$ in place
of $\epsilon$ with lemma \ref{lem:lot-auc} we get the existence of 
mechanism with menu-size complexity of $((\log m + \log H)/\epsilon)^{O(m)}$ 
whose approximation error (both additive and multiplicative) is
$\epsilon$.  The additive approximation error of the whole 
mechanism is subsumed by the multiplicative one since the optimal one has revenue of
at least $1$.  We now plug this family of low complexity mechanisms into proposition \ref{apx-C}, and obtain the required result,
in the information-theoretic sense.

To compute the the mechanism algorithmically from the sample, we first solve the linear program on the 
sample (that is of size that is polynomial in $C$,
thus exponential in $m$), obtain the optimal mechanism $M$ for the sample, and then round it to a mechanism $\tilde{M}$ that
provides the required approximation for the sample and  -- since it is of the right complexity -- also for the distribution.
\end{proof}

\subsection{Approximations for Monotone Valuations}

\label{sec:classes}
In this section, we consider a limited class of valuations and show that for distributions over this class a much smaller complexity  is needed.
The class we consider fixes an order on items, without loss of generality, the order $1,\ldots,m$. A valuation $v \in [1,H]^m$ is \emph{monotone} 
if $v_i \leq v_{i+1}$ for $i \in \set{1,\ldots,m-1}$.  Monotone valuations  are natural in contexts such as \emph{multi-unit auctions}, where $m$ identical goods are being sold, and an outcome corresponds to the number of goods allocated to the buyer. In this setting,  $v_i$ is the player's value for $i$ goods. Monotone valuations then correspond to a free disposal assumption in multi-unit auctions.  

Next we show that for the set of monotone valuations, we can find a small $\epsilon$-cover of all lotteries, which implies, using our 
``Rounding Lotteries to Rounding Auctions'' paradigm, small complexity auctions that approximate the revenue well for all
monotone valuations.  This family of auctions has menu-size complexity that is
polynomial in $m$ when $\epsilon$ and $H$ are constant, and quasi-polynomial for $H$ that is
polynomial in $m$ (and $\epsilon$ poly-logarithmic).


\begin{theorem}(Part II of theorem \ref{thm:menusize} from the introduction)
If $\D$ is supported on monotone valuations then for every 
$\epsilon>0$ there exists a menu $M$ with $C=m^{O(\frac{\log^3 H + \log^2 \epsilon^{-1}}{\eps^2})}$ entries (and with all numbers
in it given with $O(\log m + \log H + \log \epsilon^{-1})$ bits of precision) such that 
$Rev(M,\D) \ge (1-\epsilon) Rev(\D)$.   
Furthermore, this auction can be computed in $poly(C,H,\epsilon^{-1},m)$ time by sampling $poly(C,H,\epsilon^{-1},m)$ valuations from $\D$.
\end{theorem}

As in the proof of corollary \ref{partI}, using lemma \ref{lem:lot-auc} and proposition \ref{apx-C}, 
this theorem follows from the following lemma:

\begin{lemma}\label{lem:lotterycover_mono}
For every $\epsilon$, there is a set of lotteries $L$ whose size is $m^{O((\log H + \log \epsilon^{-1})/\epsilon)}$,
that $\epsilon$-covers the set of
all monotone valuations with $v:\{1...m\} \rightarrow [1,H]$.  Moreover, calculating $\tilde{x}$ from $x$ can be done efficiently.
\end{lemma}

\begin{proof}
Given lottery $x \in \Delta^m$, we define the tail-probability form $\bar{x} \in [0,1]^m$ of  $x$ as follows:  
$\bar{x}_i = \sum_{j=i}^m x_j$. We use the fact that, for a monotone valuation $v$ and a lottery $x$, the value of the lottery can be written as:
$v \cdot x = \sum_{i=1}^m (v_i - v_{i-1}) \bar{x}_i$. 

Let $Q$ be the set of real numbers containing zero and all integer powers of $(1-\epsilon)$ in the range $[\epsilon/H,1]$.
The family $\L$ consists of all lotteries $y \in \Delta_m$ who's tail probabilities are all
in $Q$: $\bar{y}_i \in Q$ for all $i$. To round a given lottery $x$ into a rounded lottery $\tilde{x}$, look at the
tail probability $\bar{x}$, round down each entry of $\bar{x}$ to a number in $Q$ obtaining a vector $\bar{y}$ and then
reconstruct a lottery from these tail probabilities $\tilde{x}_i = \bar{y}_i - \bar{y}_{i+1}$. 
Notice that this is a perfectly legal lottery since 
$\bar{y}$ is monotone (since so was $\bar{x}$) and thus $y_i \ge 0$ for all $i$ and also $\sum_i y_i = \bar{y}_1 \le \bar{x}_1 \le 1$.

Clearly $\bar{y} \le \bar{x}$, coordinate wise, and thus $\tilde{x} \cdot v \le x \cdot v$.  Now let us estimate the gap.
We separate $x \cdot v = \sum_{i=1}^m (v_i - v_{i-1}) \bar{x}_i$ to two parts, those with $\bar{x}_i > \epsilon/H$ and those
with $\bar{x}_i < \epsilon/H$.  For the first part we have $\bar{y_i} \ge (1-\epsilon) \bar{x_i}$ and so the first part is bounded
from above by $(1-\epsilon)\sum_{i=1}^m (v_i - v_{i-1}) \bar{y}_i = (1-\epsilon)\tilde{x} \cdot v$.  The second part 
is bounded from above by $\epsilon \sum_{i=1}^m (v_i - v_{i-1}) /H$ and since $v_i \le H$ it is bounded by $\epsilon$.
We thus get $x \cdot v \ge (1-\epsilon) \tilde{x} \cdot v - \epsilon$ as needed.

Finally let us calculate the size of $L$. Since $\bar{y}$ is monotone decreasing but can only take one of 
$O((\log H + \log \epsilon^{-1})/\epsilon$ values, it is determined by the few points where $\bar{y}_i \ne \bar{y}_{i-1}$.
Thus there are at most $m^{O((\log H + \log \epsilon^{-1})/\epsilon)}$.
\end{proof}

\section{Lower Bound for Auction Complexity}
\label{sec:lb}
 
\subsection{A Simple $O(\log H)$ Approximation}

Before we embark on our lower bound, we note the matching upper bound.

\begin{proposition} (proposition \ref{pro:logH} from the Introduction)
For every distribution $\D$ there exists an auction with $m$ menu-entries
(and with all numbers represented using $O(\log H)$ bits)
that extracts $\Omega(1/\log H)$ fraction of the optimal revenue 
from $\D$.  
Furthermore, this auction can be computed in polynomial time from $\poly(H)$ samples.
\end{proposition}

\begin{proof}
Let $M^p$ denote the $m$-entry menu which offers every item for a price of $p$; i.e. $M^p = \{(e_i, p) : i \in \set{1,\ldots,m}\}$, where $e_i$ is the $i$'th standard basis vector. One of the $\log H$ menus $M^1, M^2, M^4, \ldots, M^H$ is guaranteed to extract a $1/(2\log H)$ fraction 
of the expected maximum item value, $\Ex_{v \sim D} \max_i v_i$, which in turn is an upper bound on the maximum revenue. 

To make this algorithmic, observe that the revenue of each of $M^1, M^2, M^4, \ldots, M^H$ is a random variable between $0$ and $H$.  The algorithm is the obvious one: take $n=\omega(H^2 \log \log H)$ samples from $\D$, and return the menu from $M^1, M^2, M^4, \ldots, M^H$  with the greatest expected revenue on the sample. By applying the Hoeffding bound and the union bound, our estimate for the revenue of each menu $M^p$ is within an additive error of $1/2$ with high probability. Since $\Rev(\D) \geq 1$, the chosen menu satisfies the claimed bound.
\end{proof}

\subsection{The Lower Bound}

\begin{theorem} (Theorem \ref{thm:kolmogorov} from the introduction)
For every auction representation language and every $1 < H < 2^{m/400}$, there exists a distribution $\D$ on $[1..H]^m$ such that for every mechanism $M$ with complexity at most $2^{m/400}$ we have that $\Rev(M,\D) = O(1/\log H) \cdot \Rev(\D)$.
\end{theorem} 

\begin{proof}
We will construct the distribution $\D$ probabilistically.  

Our starting point will be a fixed baseline distribution $\B$ that takes an ``equal
revenue'' one-dimensional distribution and spreads it symmetrically over a random subset of the items.  
Each valuation in the support of $\B$ is specified by a set $S \subset \{1..m\}$ of size exactly $k=m/3$ and an integer scale value 
$1 \le z \le \log H$.  The valuation will give value $2^z$ for every item in $S$ and value 1 for every other item.
The probability distribution over these is induced by choosing $S$ uniformly at random among sets of size $k$ and choosing $z$ as to obtain an
``equal revenue distribution'' $Pr[z=x]=2^{-x}$.  We can view this distribution over the $v$'s as choosing  uniformly at random from a multi-set $V$
of exactly ${\binom{m}{k}} (H-1)$ valuations (for every set $S$ of size $k$ we have $H/2$ copies of a valuation with value $2$, $H/4$ copies of a valuation
with value $4$ ... and a single copy of a valuation with value $H$). The point is that due to symmetry, it can be shown that,
just like in the corresponding single dimensional case, $\Rev(\B)$ is constant (despite the expected value being $O(\log H)$).  This is proven formally
in lemma \ref{revB} below.

Taking the point of view of $\B$ being a random choice of a valuation from the multi-set $V$, we will now construct our distribution $\D$
as being a uniform choice over a random subset $V'$ of $V$, where $V'$ is of size $|V'|=K=2^{m/100}$.  We will now 
be able to provide two estimates.  On one hand, since $V'$ is a random sample from $V$, we expect that every fixed mechanism will extract approximately
the same revenue from $\D$ as from $\B$.  This can be shown to hold, w.h.p., simultaneously for {\em all} mechanisms in a 
small enough family and thus all 
mechanisms with sub-exponential complexity can only extract constant revenue.  On the other hand, as $V'$ is sparse, w.h.p. it does not contain
two valuations whose subsets have a large intersection.  This will suffice for extracting at east half of the expected value as revenue, an expected value that
is $O(\log H)$.  Lemma \ref{revDlow} below proves the former fact and lemma \ref{revDhigh} below proves the latter.
\end{proof}

\begin{lemma}\label{revB}
Let $\B$ denote the distribution above then $\Rev(\B) \le 2$.
\end{lemma}

\begin{proof}
We will prove this by reduction the the single dimensional case where Myerson's theorem can be used.
Let us define the single dimensional distribution $\G$ that gives value $2^x$ with probability $2^{-x}$
for $x \in \{1 ... \log H\}$.  The claim is that $Rev(\B) \le \Rev(\G) \le 2$.  Since $\G$ is single dimensional, the
second inequality follows from Myerson's result stating that a single price mechanism
maximizes revenue, as it is easy to verify that every possible single price gives revenue of at most 2.

To prove the first claim we will build a single-parameter mechanism for $\G$ with the same revenue as a given multidimensional one for $\B$.
Given a value $2^x$ distributed as in $\G$, our mechanism chooses a random
subset $S$ of size $k$ and constructs a valuation $v$ by combining the given single-parameter value with this set, so now $v$ is distributed according to $\B$. 
We run the mechanism that was given to us for $\B$ and when it returns  
an lottery $a=a(v)$, we sell the item in the single parameter auction with probability $\sum_{j \in S} a_j$, asking for the same payment as asked for 
the lottery $a$ in the $\B$-auction.  Now notice that the same outcome that maximizes utility for $v$ also maximizes utility for the single parameter buyer.
\end{proof}

\begin{lemma}\label{revDlow}
Let $\M$ be the set of mechanisms with complexity at most $2^{m/400}$ and choose the distribution $\D$ 
as described above then, w.h.p., $\Rev(\M,\D) \le 3]$.
\end{lemma}

\begin{proof}
Since there are at most $2^{2^{m/400}}$ mechanisms in $\M$, this follows from the following lemma and the union bound.
\end{proof}

\begin{lemma}
Fix some mechanism $M$ and choose the distribution $\D$ 
as described above then $Pr[\Rev(M,\D) > 3] \le exp(-K/H^2) \le 2^{-2^{m/300}}$.
\end{lemma}

\begin{proof}
Let $r_M(v)$ denote the revenue that mechanism $M$ extracts on valuation $v$.
By definition $Rev(M,\B) = E_{v \in V} r_M(v)$ while $Rev(M,\D) = E_{v \in V'} r_M(v)$ (where the distribution is uniform
over the multi-sets $V$ and $V'$ respectively).  Lemma \ref{revB} bounded the former: $Rev(M,\B) \le 2$.  
Since we are choosing $V'$ to be a 
random multi-set of size $K$ and since for all $v$ we have $0 \le r_M(v) \le H$ then we can use Chernoff bounds
to bound the probability that the expectation of $r_M(v)$ over the
sample $V'$ is larger than its expectation over the population $V$ to be
$Pr [|E_{v \in V'} r_M(v) - E_{v \in V} r_M(v)| > 1] \le exp(-K/H^2)$.
\end{proof}

\begin{lemma} \label{revDhigh}
Choose the distribution $\D$ as described above then, w.h.p, $\Rev(\D) \ge \log H / 2$.
\end{lemma}

\begin{proof}
We will use the following property that holds, w.h.p., for $V'$: for every two different valuations in $V'$ the sets of items $S$ and $T$ 
associated with them satisfy $|S \cap T| < m/6$.  The reason that this property holds is
that for any fixed $T$, since $S$ is a random set of size $m/3$ the probability that
$|S \cap T| / |T| \ge 1/2$ is $exp(-|T|) \le 2^{-m/40}$.  Now we can take a union bound over all $K^2 = 2^{m/50}$
possible pairs of $S$ and $T$. 

Using this property of $\D$ here is a mechanism that extracts as revenue at least half of the expected value of $v$, i.e. at least $(\log H)/2$ revenue:
we have a menu entry for each element $v \in V'$.  For
$v$ that gives value $2^z$ to the set $S$, this entry will
offer every item in $S$ with probability $1/|S| = 3/m$, and will ask for payment of $2^{z-1}$ for this lottery.
Clearly if $v$ chooses this entry it gets net utility of exactly $2^{z-1}$, we need to show that the net utility from any
other menu entry is less than this.  Observe that $v$'s value from a menu entry that corresponds
to a set $T$ is exactly $2^z \cdot |S \cap T| / |T|$ which due to our property is bounded from above by $2^{z-1}$.
\end{proof}

\section{Computational Complexity}
\label{sec:computational}

In this section, we examine the computational complexity of the algorithmic task associated with Proposition \ref{apx-C} when valuations lie in $[1,H]^m$. We restrict our attention to the menu-complexity model. The computational bottleneck is Step 2 of the algorithm, which computes a $C$-menu maximizing revenue for the uniform distribution over samples. 
Specifically,  step 2 of the algorithm requires the solution of the following optimization problem $MAXREV$. An instance of $MAXREV$ is given by an integer $C$ and a sample $X=\set{v_1,\ldots,v_n} \sse [1,H]^m$. Feasible solutions of $MAXREV$ are menus with at most $C$ entries, and the objective is to maximize revenue for a buyer drawn uniformly from $X$. We leave essentially open the exact computational complexity of approximating $MAXREV$, yet make some progress by showing the problem APX-hard.

\begin{theorem}\thmcontinues{thm:maxrev_nphard}
$MAXREV(X, C)$ in NP-hard to approximate to within any factor better than $1-\frac{1}{e} \frac{H-1}{H}$, even for valuations in $\set{1,H}^m$.
\end{theorem}

\begin{proof}
We reduce from a promise problem related to the NP-hard optimization problem \emph{max cover}. 
For convenience, we use the equivalent \emph{hitting set} formulation of max cover. 
The input to this problem is a family $\S=\set{S_1,\ldots,S_n}$ of subsets of $[m]$, 
and an integer $k$, and the outputs is a ``hitting set'' $T \sse [m]$ of size at most $k$ 
maximizing the number of sets $S \in \S$ with which $T$ has a non-empty 
intersection --- we say those sets $S$ are ``hit'' by $T$. We use the fact that it is 
NP-hard to distinguish between instances of hitting set in which the optimal solution hits all 
sets in $\S$, and instances in which the optimal solution hits less than a $1-\frac{1}{e} + \epsilon$ fraction 
of the sets in $\S$, for any constant $\epsilon>0$ (see Feige \cite{feige_setcover}).

Given an instance $(\S,k)$ of hitting set, we produce an instance $(X,C)$ of $MAXREV$ as follows.  
We let $C=k$, and for each $S_i \in \S$ we include a valuation $v_i \in X$ such that $v_i(j) = H$ 
for $j \in S_i$, and $v_i(j)=1$ otherwise. If there is a hitting set $T$ of size $k$ which hits 
every $S_i \in \S$, then the item-pricing $C$-menu $\set{(e_j,H) : j \in T}$, which prices every 
item  $ j \in  T$ at $H$, generates a revenue of $H$ from every valuation $v_i \in X$. 
On the other hand, we show that if there is a $C$-menu with average revenue at 
least $R=H-{(\frac{1}{e} - \eps)(H-1)}$ over $X$, then there is a hitting set of size $k$ hitting at 
least a $\frac{R-1}{H-1} = 1 - \frac{1}{e} + \eps$ fraction of the sets in $\S$. 
Consider such a $C$-menu $M=\set{(x_1,p_1),\ldots,(x_C,p_C)}$, and draw an item $j_t$ from 
each lottery $x_t$ in $M$.\footnote{Since our lotteries are partial --- i.e. $\sum_j x_t(j) \leq 1$ --- some of 
the items $j_t$ may be the ``null'' item.}  Let $T= \set{j_1,\ldots,j_C}$ be the resulting 
random hitting set of size $\leq C=k$.  It suffices to show that $T$ hits at least an $\frac{R-1}{H-1}$ 
fraction of the sets in $\S$ in expectation. The following calculation completes the proof.

\begin{align*}
 R &=  \avg_{i=1}^n \Rev(M,v_i) & &\\ 
 &\leq \avg_{i=1}^n \max_{t=1}^C v_i \cdot x_t & & \mbox{(\ by individual rationality \ )} \\
 &= \avg_{i=1}^n \max_{t=1}^C \left[ H \cdot x_t(S_i) + 1 \cdot x_t([m] \sm S_i) \right]& & \mbox{(\ $x_t(S)$ denotes $\sum_{j \in S} x_t(j)$\ )} \\
&\leq \avg_{i=1}^n \max_{t=1}^C \left[ H \cdot x_t(S_i) + 1 - x_t(S_i) \right]& & \mbox{(\ because $\sum_j x_t(j) \leq 1$ \ )} \\
&= 1+ (H-1) \avg_{i=1}^n \max_{t=1}^C  x_t(S_i)& &\\  
&= 1+ (H-1) \avg_{i=1}^n \max_{t=1}^C  \Pr [ j_t \in S_i]& & \mbox{(\ because $j_t \sim x_t$ \ )} \\
&\leq 1+ (H-1) \avg_{i=1}^n  \Pr [ T \intersect S_i \neq \emptyset]  
\end{align*}
\end{proof}

The reader might have noticed that step 2 of the `Sample-and-Optimize' algorithm template, requiring the solution of an 
instance of $MAXREV$, is too restrictive as stated. An auctioneer may constrain the complexity of his sought mechanism 
either because it is believed that such a mechanism is approximately optimal \footnote{For some structured 
distributions (but still with exponential-size support, e.g. independent valuations across items), such belief can be 
formally proved.}, or because computational and/or practical considerations limit the auctioneer to only ``simple'' mechanisms. 
In both cases, it is perhaps more natural 
to seek a \emph{bicriteria guarantee}: a mechanism of complexity polynomial in $C$, $m$, and $\log H$ which nevertheless 
approximates the revenue of the best mechanism of complexity $C$. 

To illustrate this idea, consider the following variant of the `Sample-and-Optimize' algorithm with step $2$ 
replaced by its \emph{bicriteria} version:

\begin{enumerate}
\item Sample $t= poly_1(C, m, \eps^{-1}, H)$ samples from $\D$;
\item Find an auction of complexity at most $poly_2(C, m, \eps^{-1}, \log H)$ that approximates the optimal auction of complexity $C$ on the $t$ samples;
\end{enumerate}

Using the same idea in the proof of Prop \ref{apx-C}, we can see that, in order to 
avoid over-fitting for auctions of complexity at most $poly_2(C, m, \eps^{-1}, \log H)$, a sample 
size of $poly_1(C, m, \eps^{-1}, H)$ suffices, as long as $poly_1$ is a somewhat larger polynomial than $poly_2$. 
So with an $\alpha$-approximation for the bicriteria $MAXREV$ problem, we have an efficient $\alpha$-approximation 
algorithm in the \emph{black-box} model.   We leave the approximability of this bicriteria variant of $MAXREV$ as an open question.

{
\bibliography{pricing}
\bibliographystyle{plain}
}

\appendix






  

\end{document}